# Fabrication and Electro-optic Properties of MWCNT Driven Novel Electroluminescent Lamp

*D. Haranath[*], Sonal Sahai, Savvi Mishra, M. Husain[#] and Virendra Shanker*

[*]CSIR-National Physical Laboratory, Dr K S Krishnan Road, New Delhi – 110 012, INDIA

[#]Centre for Nanoscience and Nanotechnology, Jamia Millia Islamia, New Delhi- 110 025, INDIA

**Abstract**

We present a novel, cost-effective and facile technique, wherein multi-walled carbon nano-tubes (CNTs) were used to transform a photoluminescent material to exhibit stable and efficient electroluminescence (EL) at low-voltages. As a case study, a commercially available ZnS:Cu phosphor (P-22G having a quantum yield of 65±5%) was combined with a very low (~0.01 wt.%) concentration of CNTs dispersed in ethanol and its alternating current driven electroluminescence (AC-EL) is demonstrated. The role of CNTs has been understood as a local electric field enhancer and facilitator in the hot carrier injection inside the ZnS crystal to produce EL in the hybrid material. The mechanism of EL is discussed using an internal field emission model, intra-CNT impact excitation and the recombination of electrons and holes through the impurity states.

Keywords: Electroluminescence, photoluminescence, hybrid materials, carbon nanotubes.

[*] Corresponding author E-mail: haranath@nplindia.ernet.in (D. HARANATH)

1. **Introduction**

Alternating current electroluminescent (AC-EL) lamps already occupy a segment of the high-resolution, flat panel display market. AC-EL displays in thin-film form are robust, possess long lifetimes, and offer high luminance with relatively low power consumption [1]. However, AC-EL lamps in thick film form have attracted considerable interest for illumination and display applications, due to their substrate flexibility, simple fabrication process of large area panels and low cost fabrication avoiding requirement of vacuum processing [2-4]. Such AC-EL lamps consist of a phosphor layer, e.g. copper-doped zinc sulfide (ZnS:Cu), vertically sandwiched between two insulators that are contacted by electrodes. When a sufficiently high voltage is applied across the electrodes, electron charge carriers are injected into the conduction band of the phosphor, where they are accelerated to high energies by the field and energy transfer takes place via impact excitation of the luminescent centers and finally radiative relaxation results the EL [5, 6].

Despite the simple fabrication processes of EL thick film type lamps, their applications have been confined to backlight units in cellular phones due to deficient performance as low brightness and efficiency [7]. In spite of considerable efforts in improving the performance of AC-EL lamps, the fabrication of simple and stable structure operated at low voltage and low frequency is still required and has been the greatest challenge to the scientific community [8, 9]. This paper reports the triggering of EL in a non-EL material with the use of commonly available multi-walled carbon nano tubes (CNTs) and it is anticipated that use of CNTs can improve the EL performance of AC-EL lamps. The mechanism to understand this unusual EL phenomenon has also been successfully explained.

The most investigated EL phosphor for conventional AC-EL lamps is copper-doped zinc sulfide (ZnS:Cu) wherein high doping of Cu ~1000 ppm is present [10, 11]. It is well-known that excess doping of copper is a pre-requisite for the phosphor to exhibit EL. This has dual functionality of primarily forming electrically conducting $Cu_xS$ needles inside the ZnS grain and secondly creating the required luminescent centers in the ZnS host lattice. When high voltage is applied, the $Cu_xS$ needles provide enhanced local electric field in the ZnS grains [12, 13], thus producing EL, as represented in figure 1(a). However, the EL lamps fabricated out of them has a serious operational problem of prolonged usage. With time, the sulfur diffuses into the lattice, resulting in the awful performance and drastic reduction of its lifetime.

In the current research, after rigorous experimentation and the development of many strategies of device fabrication, a novel idea of replacing $Cu_xS$ conducting needles by commonly available multi-walled carbon nano-tubes (CNTs) have been explored to engender smart EL. It is well-known that CNTs are stable and outstanding field enhancing materials due to their unique properties of high aspect ratio [14], excellent electrical [15] and thermal conductivities [16], mechanical strength [17] and chemical inertness. This makes them an ideal substitute for $Cu_xS$ conducting needles as shown in figure 1 (b). The work was carried out in this direction, with an understanding that, when $Cu_xS$ needles are replaced by CNTs, a high electric field region naming *hot spot* would necessarily be formed between the nearest placed CNTs within the ZnS grain, which can cause EL emission. For the purpose, CNT based AC-EL lamp was developed using ZnS:Cu phosphor to check the concept and smart EL was found to be triggered at very low operating voltages. (See the Supporting Information for the video clip).

## 2. Experimental

*2.1 Sample preparation*

A novel, cost-effective and facile technique has been suggested, wherein a non-electroluminescent (non-EL) material is being converted to exhibit stable and efficient EL at low-voltages using CNTs. Commercial cathodoluminescent phosphor, ZnS:Cu (P-22G with quantum yield 65±5%) having minimal (~50 ppm) Cu doping, procured from M/s Samtel Color Lab., India, was used as starting material. The multi-walled carbon nano-tubes (CNTs) were grown by chemical vapor deposition technique in the Carbon section of CSIR-NPL, India. The weighing of lower amounts of CNTs with precision is a tedious job. Hence, we have dispersed the known weight of CNT in ethyl alcohol and ultra-sonicated rigorously for 2 h and made a suspension. This was further added to ZnS:Cu phosphor and allowed to dry at room temperature (~25°C). After complete drying the mixture was blended mechanically was annealed at 450°C under the inert $N_2$ atmosphere. The amount of CNT addition is found to be highly critical in our experiments. An admissible range to exhibit EL was only 0.005-0.03 wt%. However, ~0.01 wt% of CNT was found to be optimum, which showed stable and bright EL from the ZnS/CNT hybrid material. Another important finding in our experiment is the addition of ~5 wt% of potassium bromide (KBr) as a chemical "additive", which was mixed with the ZnS/CNT blend before annealing in order to diffuse CNTs into the ZnS lattice more effectively. This further helps in creating *hot spots* inside the ZnS grain when there is an applied electric field, triggering smart EL as shown in figure 1. Afterwards, blends of ZnS and varied concentrations of CNTs were prepared and annealed in the temperature range of 400°C to 600°C for 1 to 3 h and checked for their EL response.

Stable and significantly good EL was found to be exhibited by the sample annealed at/above 500°C for 2 h under inert $N_2$ atmosphere. Hence, 500°C was fixed as a suitable annealing temperature for ZnS/CNT blend for further studies. The ZnS/CNT samples annealed at 500°C, so obtained were characterized for various parameters as described in succeeding sections.

*2.2. Sample characterization*

For phase identification, the structural characterization was performed using XRD (Rigaku: MiniFlex, Cu Kα; λ=1.54 Å). The surface morphology and microstructural characterization was carried out by scanning electron microscopy (SEM, model number LEO 440). Room temperature steady-state luminescence characterization was done using a Perkin Elmer Luminescence Spectrometer (Model No. LS-55). The electro-optical characterizations of the samples have been performed using indigenously made set-up shown in figure 5 and discussed in the latter sections.

A prototype EL lamp of the size 15 x 40 mm² was fabricated at room temperature by using ZnS:Cu phosphor and MWCNT hybrid material by conventional spreading technique on indium tin oxide (ITO) coated conducting glass and screen printing of aluminum electrode on the top. This enabled us to complete the final EL lamp structure.

3. **Results and Discussion**

Figure 2 shows the photoluminescence (PL) spectrum of commercial ZnS:Cu (P-22G) phosphor as received from M/s Samtel Color Lab., India. It showed a bright green PL emission (quantum yield 65 ± 5%) having peak maximum at 535 nm, when observed under an excitation wavelength of 335 nm (3.7 eV). The doped copper ($Cu^+$) ions form the required acceptor levels at ~1.29 eV above the valence band of the ZnS lattice [18]. Inset of figure 2 shows the energy band scheme for copper doped ZnS phosphor depicting its well-known band gap ~3.6 eV. This implies that the excitation energies higher than the band gap could very well be absorbed by the ZnS crystal. A radiative recombination of electrons and holes occur at the $Cu^+$ acceptor level emits the desired green (535 nm) light. It is important to note that ZnS:Cu phosphor in its as received form did not exhibit EL in the range 10-1000 $V_{pp}$ AC, which is due to insufficient doping of copper present in the crystal system. This may be attributed to the lack of conducting paths induced by $Cu_xS$ needles inside the ZnS grain.

The phase analysis of the ZnS/CNT hybrid materials annealed at different temperatures has been shown in figure 3. It is observed that the ZnS phosphor used in the current study exhibits cubic nature with sphalerite structure. The presence of planes B1-[111]; B2-[200]; B3-[220]; B4-[311]; B5-[400]; B6-[331] confirm zinc blend structure [JCPDS card no: 005-0566]. The peaks have been identified and indicated in the XRD pattern shown in figure 3. The XRD profiles did not show the peaks related to CNTs as they are present in nominal (~0.01 wt%) amounts in any of the samples treated for high temperature annealing cycles. This substantiates the fact that the original phase of the ZnS phosphor is **retained** even after annealing treatment.

The morphological studies of the ZnS:Cu phosphor, CNT, their mechanical blend and the annealed hybrid material were performed using a scanning electron microscope (SEM) and are shown in figure 4(a-d). From the SEM image (figure 4(a)), the average particle size of ZnS:Cu phosphor has been estimated to be ~5 μm. The SEM of multi-walled CNTs used in the experiment shows that the CNTs have high aspect ratio with an average diameter of ~100 nm (figure 4(b)). The corresponding ZnS particles and CNTs have been marked in the SEM micrographs (figure 4(c) and 4(d)) for easy visualization. The SEM image (figure 4(d)) of the representative ZnS/CNT hybrid material annealed at 500°C clearly shows that the CNTs have been diffused appropriately into ZnS particles due to heat treatment. The same has been marked in the images for convenience.

A schematic diagram of the EL device fabricated in the current study is shown in figure 5. A series of ZnS/CNT hybrid materials were prepared under various conditions of annealing and CNT concentrations. A prototype device for each hybrid material was fabricated and tested using the above scheme. Prior to the fabrication, dielectric constant of the epoxy, used as dielectric medium in the device, was calculated by measuring capacitance using LCR meter and was found to $\varepsilon_r$~4.99. In a typical fabrication process, the ZnS/CNT hybrid phosphor was mixed well with dielectric medium ($\varepsilon_r$~4.99) and coated onto the conducting surface of transparent ITO glass using conventional spreading technique. The thickness of the phosphor layer was approximately around 40 μm. After drying at 50°C, another dielectric layer (thickness ~10 μm) was coated over the phosphor layer as a buffer and to avoid shorting of the device. Finally, an aluminum (Al) metal layer was screen printed at the top, which served as back-electrode. The electrical contacts were drawn out from ITO coated glass and Al metal. The AC voltage with varied amplitude and frequency was applied to the device through the contacts. The observed EL was measured using a photomultiplier tube (PMT). The EL measurements for the entire delegate devices were performed under similar conditions such as, distance of sample from the PMT and the active area of the device exposed to PMT, ambient conditions of temperature, atmospheric pressure, relative humidity etc. for relative comparison.

The dependence of EL brightness over applied AC voltage (B-V curve) for hybrid material with varied CNT concentrations in the range 0 to 0.2 wt% have been shown in figure 6. For the sample without CNT addition, EL emission was not observed up to 1000 $V_{pp}$ AC. However, with an introduction of CNT in a nominal quantity, EL was observed for very low applied AC voltage. In other words, the threshold voltages to drive the EL device decreased enormously from unknown values to <100 V AC by addition of CNTs. The measurable EL emission was observed for a sample having the CNT concentration of ~0.01 wt% and in this case, the threshold voltage reduced to ~50 $V_{pp}$ AC. It is worth mentioning here that ~0.01 wt% concentration of CNTs is found to be the optimum to exhibit stable and maximum EL brightness as shown in figure 6. However, when the CNT concentration was >0.2 wt% in ZnS/CNT hybrid material, the device burned-out with sparking due to large conductivity. Thus, the concentration of CNT in ZnS/CNT hybrid material is critical for proper functioning of the EL device.

Keeping this in view and above mentioned result, the concentration of CNT in the hybrid material was fixed at ~0.01 wt% for further studies. Inset of figure 6 shows the EL spectrum recorded for ZnS/CNT hybrid material having ~0.01 wt% CNT, indicating emission peak centered at 528 nm for 350 $V_{pp}$ and 2.65 kHz frequency. It is important to note here that the EL spectrum has been observed to blue shifted by ~7 nm when compared to the PL spectrum of ZnS:Cu phosphor as shown in figure 2.

All the ZnS/CNT hybrid materials that show EL emission was found to obey the following empirical relation:

$$B = \begin{cases} B_0 \exp\left(-\dfrac{b}{V^{0.5}}\right), & 0 \leq V < 260 \text{ V} \\ B_0 \exp(V), & V \geq 260 \text{ V} \end{cases}$$

where, B is the brightness, V is the applied voltage, $B_0$ and *b* are the constants determined by the particle size of the phosphors, structure of the device and the exciting bias conditions [10]. The B-V relation in the range 0<V<260 V implies well-known phenomenon of EL in ZnS phosphors [19]. Usually the EL brightness of the device increases with an increase in the applied AC voltage and attains a saturation value. A typical increase of applied voltages beyond brightness saturation leads to decrease in capacitance value and hence, shorting of the device. The brightness-voltage relation could be explained using the above said equation. In the first part, 0<V<260 V, EL mechanism is responsible for the well-known process of tunneling of electrons as well as injection of hot electrons in the ZnS crystal [20]. Whereas, the presence of CNTs seems to have modified the B-V relationship as represented by second part of the equation, V≥260 V, which may be due to efficient intra-CNT impact excitation by hot carriers and the development of *hot spots* inside the ZnS grains [21].

The measurement of EL brightness with respect to frequency is another important feature of EL study. The variation in EL brightness with frequency has been shown in figure 7. Emission intensity was observed to increase with AC frequency initially, reaches to maximum at a frequency around 2.65 kHz and start deteriorating afterwards. A possible explanation for this may be as follows: As the number of cycles per second increases, the energy supplied to the device also increases, reinforce the EL brightness. When frequency is further increased beyond the optimum, the extra energy starts dissipating as heat, leading to decrease in brightness and finally damage of EL device [22].

Inset of figure 7 shows a marginal (~12 nm) variation of emission wavelength flaunted by ZnS/CNT hybrid material under various applied AC voltages. With the increase in voltage EL emission peak experiences blue shift. This may be attributed to the phonon assisted EL emission by CNTs [23].

EL brightness of ZnS/CNT hybrid material has been found to be influenced by the exciting input waveforms as shown in figure 8. The sinusoidal wave delivered the optimum EL brightness compared to square and saw-tooth waves. The reason for the same could be explained in terms of distribution of voltages in different waveforms thus leading to the injection of hot electrons into the device. The injection of hot electrons and extraction of hot holes in sinusoidal wave is gradual which in turn excites luminescent centers uniformly, thereby increasing the radiative recombination invigorating to an efficient EL. Although, saw-tooth wave also injects hot carriers gradually, the abrupt jump from positive to negative voltage reduces the excitation as well as the recombination rates, leading to decrease in the EL brightness. The rapid jump from negative to positive (or vice versa) in square wave causes a massive charge injection into the emissive layer, resulting in an unbalanced situation between much higher densities of charge of both polarities. This behaviour also leads to a further decrease in the EL brightness [24].

AC-EL device using ZnS/CNT hybrid material has been fabricated and analyzed for the first time, for photo- and electroluminescence and corresponding electro-optical properties; such as brightness-voltage, brightness-frequency and brightness-waveforms. It was found that conducting paths induced inside ZnS phosphor for effective electron and/or hole-transport by the use of CNT was the main cause of realizing low voltage AC-EL device.

Figure 9 (a) shows the photograph of the prototype device depicting two spots related to intrinsic ZnS:Cu phosphor and ZnS/CNT hybrid material, respectively, operated at a threshold AC voltage ~50 $V_{pp}$ and 2.65 kHz. It can be clearly seen from the photograph that the spot corresponding to intrinsic ZnS:Cu phosphor does not show EL at all while a very bright and stable green EL emission is seen from the ZnS/CNT hybrid material. The same has been marked in the photograph. Thus it can be concluded that a non-EL material has been translated to exhibit EL by the introduction of nominal (~0.01 wt%) concentration of CNTs.

The mechanism of EL is highly important to understand the working of the device. Figure 9 (b) illustrates the energy band diagram for the ZnS/CNT hybrid material. When the voltage is applied across the device, CNTs produce high carrier mobility and local electric fields which lowers the energy barrier at the contact aluminum electrode facilitating efficient tunneling of electrons into the device. This process is known as *internal field emission* [21]. Due to the enhancement in local electric field at the ends of CNTs, some *hot spots* are being created between the nearest CNTs, within the ZnS grain as shown in figure 1 (b). These region of high electric fields, promote charge carriers to be injected into ZnS:Cu phosphor, which get accelerated to higher energy and transfer their ballistic energy to luminescent centers. As a result excitation of luminescent centre takes place and finally relaxation gives the EL emission.

4. **Conclusions**

The EL from the non-electroluminescent ZnS:Cu was successfully achieved with the introduction of CNT. This study was a pilot study and showed productively that a bright, stable and low threshold EL can be obtained from ZnS/CNT hybrid material by a cost-effective and easy fabrication process. The electro-optical properties such as brightness-voltage, brightness-frequency and brightness-waveforms were also presented. The phenomenon observed is new and has not been reported earlier. It is anticipated that the idea will be very interesting and robust in the field of AC-EL.

**Acknowledgements**

The authors (DH, SS and SM) wish to gratefully acknowledge the Department of Science and Technology, Government of India for the financial support under the schemes # SR/WOS-A/PS-03/2009 and SR/FTP/PS-012/2010 to carry out the above research work.

[1] Y. A. Ono, in *Electroluminescent Displays*; World Scientific:River Edge, NJ, **1995**.

[2] J. Y. Kim, S. H. Park, T. Jeong, M. J. Bae, S. Song, J. Lee, et al. *IEEE Trans Electron Dev* **2010**, 57(6), 1470.

[3] T. Satoh, T. Nakatsuta, K. Tsuruya, Y. Tabata, T. Tamura, Y. Ichikawa, et al. *J. Mater Sci. Mater Electron* **2007**, 18, S239.

[4] M. Warkentin, F. Bridges, S. A. Carter, M. Anderson, *Phy. Rev. B* **2007**, 75, 075301.

[5] J. P. Keir, J. F. Wager, *Annu. Rev . Mater. Sci.* **1997**, *27*, 223.

[6] M. Dur, S.M. Goodnick, S. S. Pennathur, J. F. Wager, M. Reigrotzki, R. Redmer, *J. Appl. Phys.* **1998**, *83*, 3176.

[7] G. Sharma, S. D. Han, J. D. Kim, S. P. Khatkar, Y. W. Rhee, *Ma.t Sci. Eng. B* **2006**, 131, 271.

[8] K. Hirabayashi, H. Kozawaguchi, B. Tsujiyama, *J. Electrochem. Soc.* **1983,** 130, 2259.

[9] A. P. Alivisatos, *Science* **1996,** 271, 933.

[10] H. Chander, V. Shanker, D. Haranath, S. Dudeja, P. Sharma, *Mat. Res. Bull.* **2003**, 38, 279-288.

[11] N. Jing-hua, H. Rui-nian, L. Wen-lian, L. Ming-tao, Y. Tian-zhi, *J. Phys. D: Appl. Phys.* **2006**, 39, 2357-2360.

[12] K. Maeda, *J. Phys. Soc. Japan* **1958**, 13, 1351.

[13] A. G. Fischer, *J. Electrochem Soc.* **1963**, 110, 733.

[14] S. Iijima, *Nature (London)* **1991**, 354,56.

[15] H. Dai, E.W. Wong, C. M. Lieber, *Science* **1996**, 272, 523.

[16] S. Berber, Y. K. Kwon, D. Tomanek, *Phys. Rev. Lett.* **2000**, 84, 4613.

[17] M. M. J. Treacy, T. W. Ebbesen, J. M. Gibson, *Nature (London)* **1996**, 381, 678.

[18] S. Shigeo, M.Y. William, *Phosphor Handbook*, first ed., CRC Press, USA, **1999**.

[19] A. G. Fischer, *J. Electrochem Soc*. **1962**,109, 1043.

[20] P. Zalm, *Philips Res.*, Rep No. 11 **1956**, 353, 417.

[21] P. R. Thornton, in *The Physics of Electroluminescent Lamps*, (Eds: E. J. Burge), E. & F. N. Spon Limited, London, **1967**.

[22] K. Miyashita, M. Wada, T. Takahashi, *Japanese Journal of Applied Physics* **1964**, 3(1), 1.

[23] S. Essig *et al., Nano Letters* **2010**, 10, 1589.

[24] F. Baudoin, D. H. Mills, P. L. Lewin, S. L. Roy, G. Teyssedre, C. Laurent, *J. Phys. D: Appl.Phys*. **2011**, 44, 165402.

**Figures with captions:**

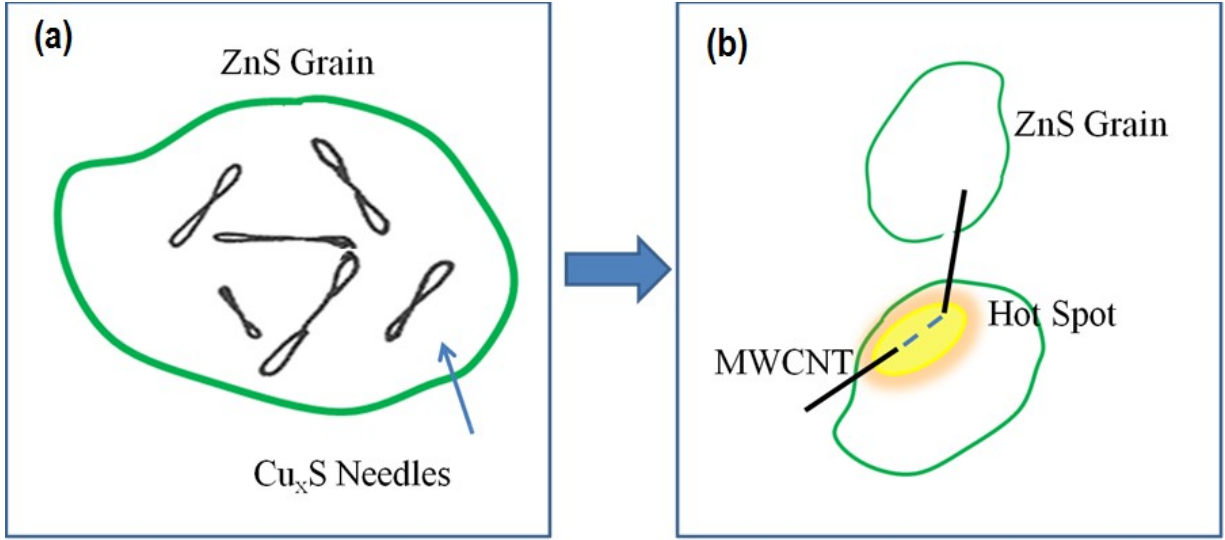

**Figure 1.** (a) Schematic showing the formation of $Cu_xS$ needles in Cu doped-ZnS grains, which is used in conventional EL lamps and (b) appearance of *hot spot* in CNT based EL lamps (current work) due to the tunnelling of electrons between nearest CNTs when the field is applied.

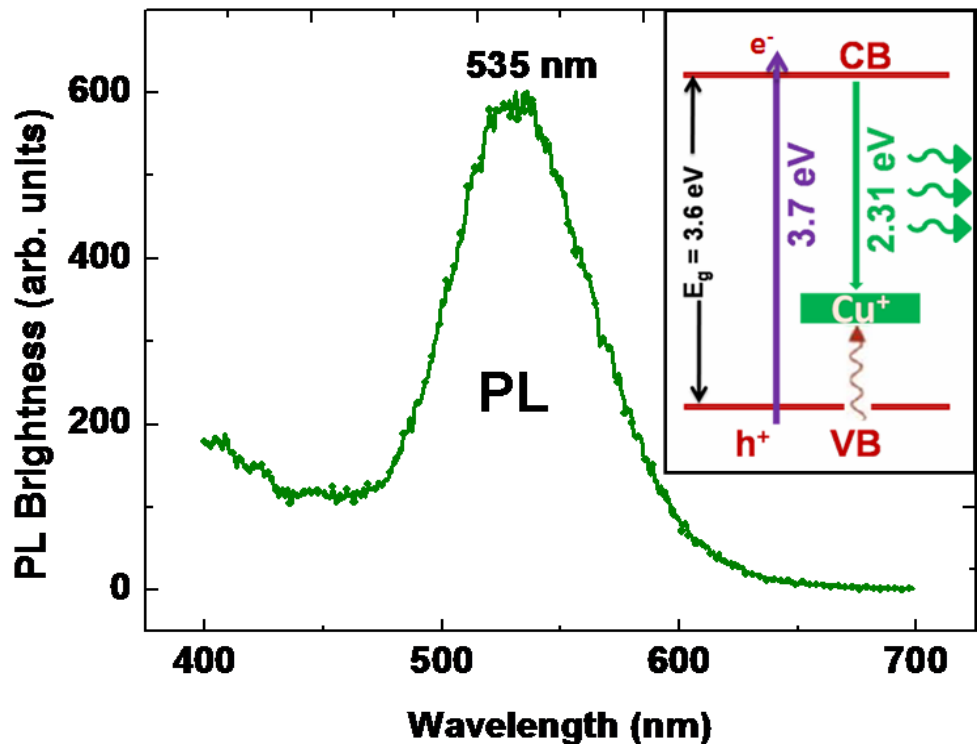

**Figure 2.** The photoluminescence spectrum of commercial ZnS:Cu (P-22G with quantum yield 65±5%) phosphor shows an emission peak at 535 nm at excitation wavelength of 335 nm (3.7 eV). Inset shows the schematic energy level diagram of possible electronic transitions corresponding to absorption and radiative recombination at $Cu^+$ site leading to green emission at 535 nm (2.31 eV).

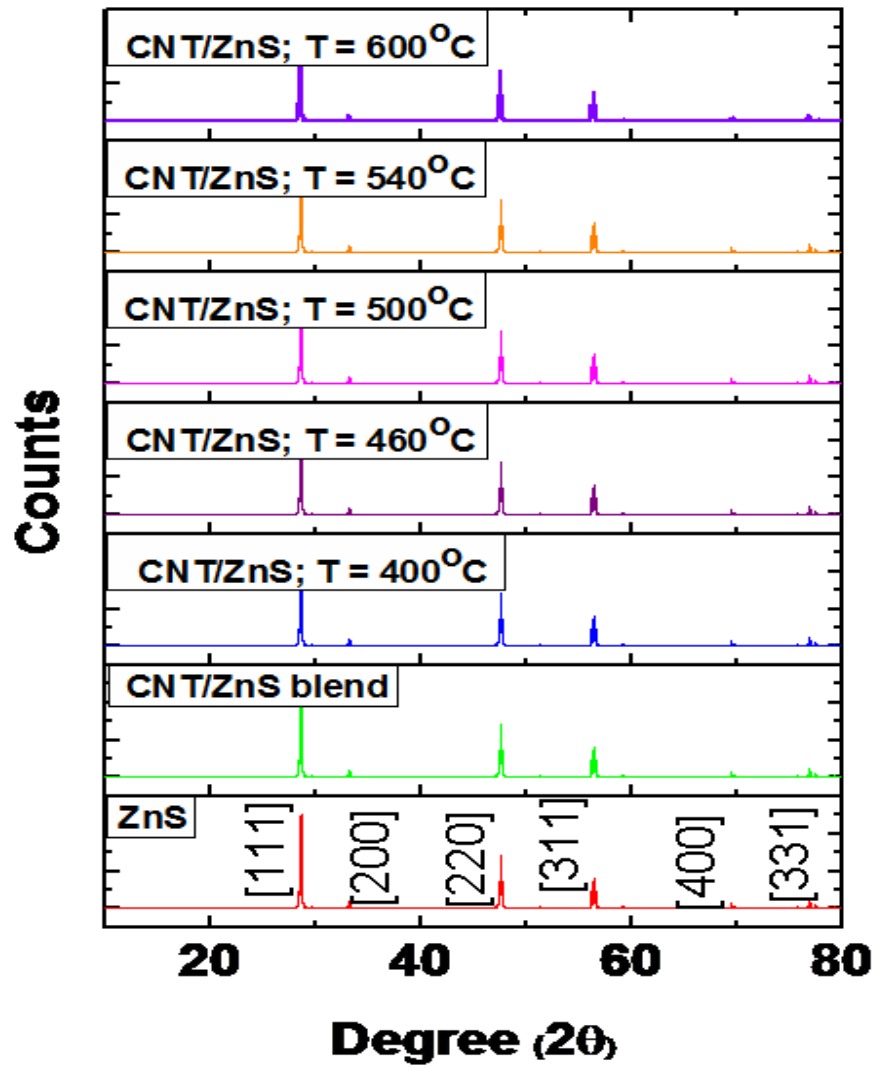

Figure 3. X-ray diffraction patterns of ZnS/MWCNT hybrid materials annealed at different temperatures ranging room-temperature to 600°C under inert $N_2$ atmosphere depicts almost no change in the crystal structure of ZnS.

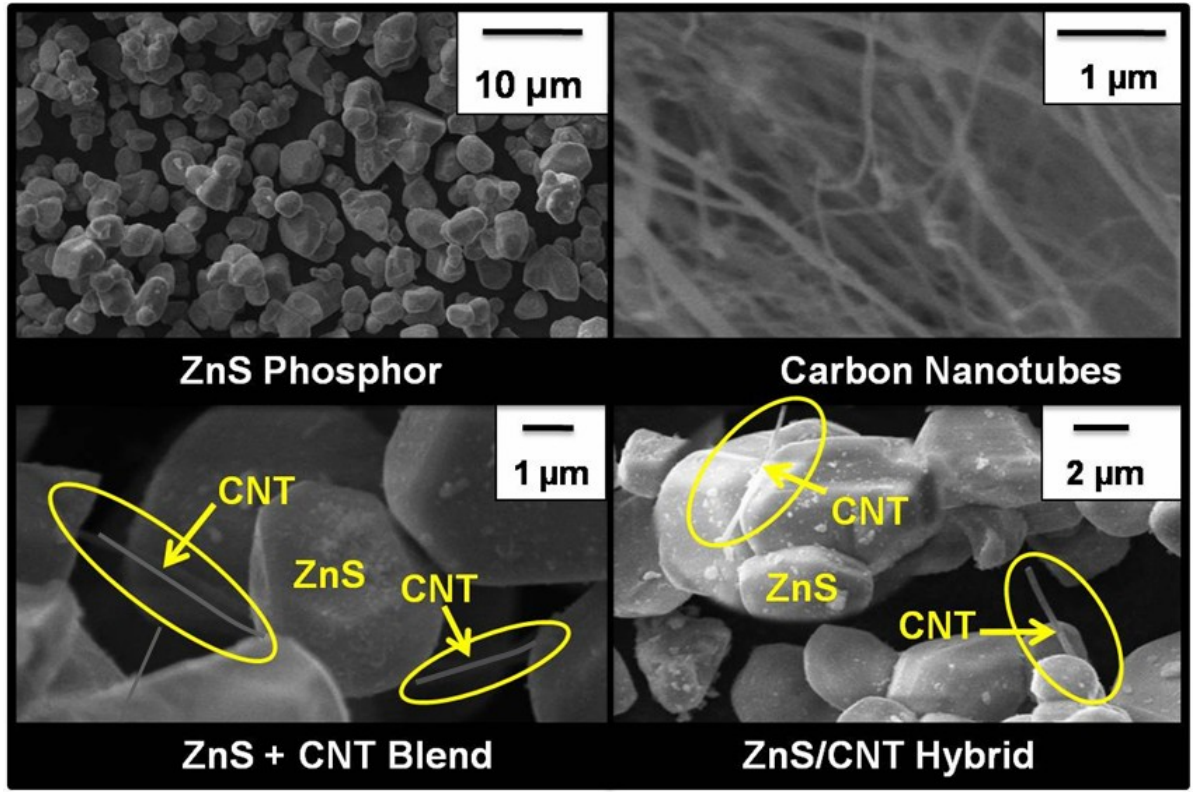

**Figure 4.** SEM images show morphology of ZnS phosphor, carbon nanotubes, ZnS+MWCNT blend and ZnS/MWCNT hybrid material after annealing.

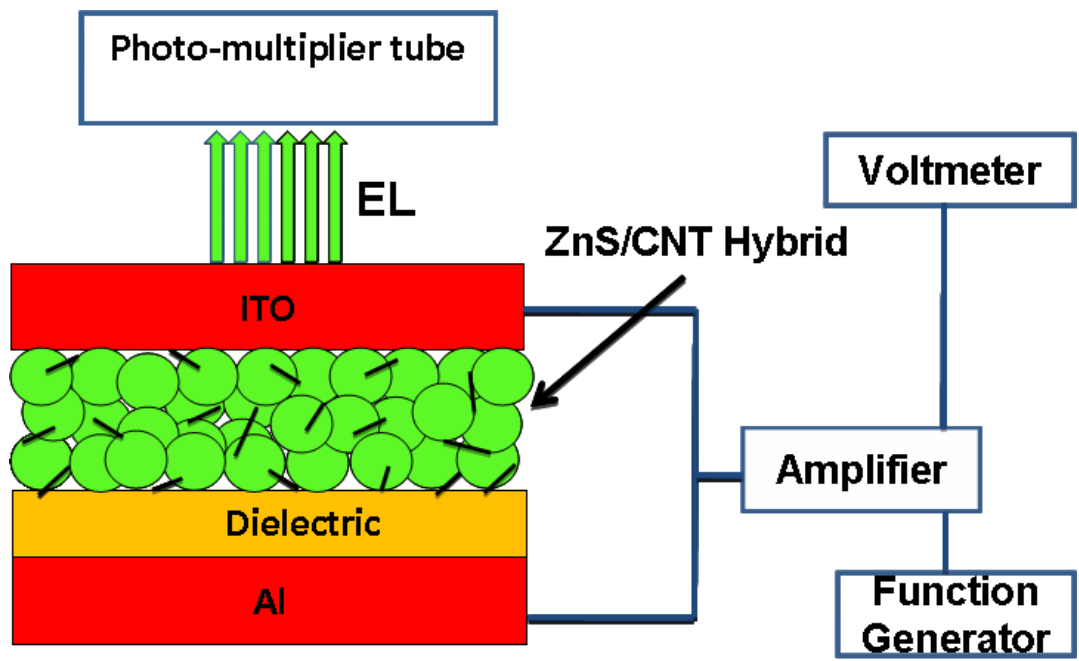

**Figure 5.** A schematic of the ZnS/MWCNT hybrid material-based AC-EL smart lamp structure.

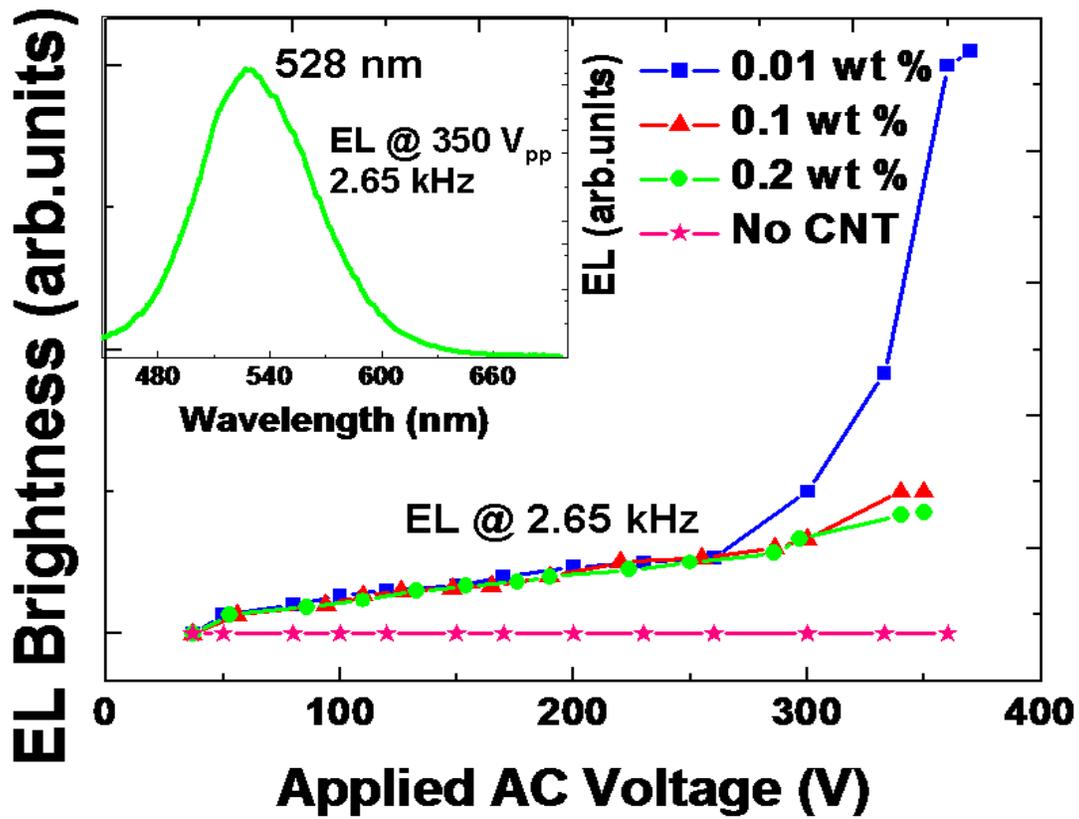

**Figure 6.** Plot of the electroluminescent (EL) intensity versus the applied AC voltage across the lamp structure. Inset of the figure shows the EL spectrum of the lamp at a given AC voltage of 350 $V_{pp}$ at 2.65 kHz.

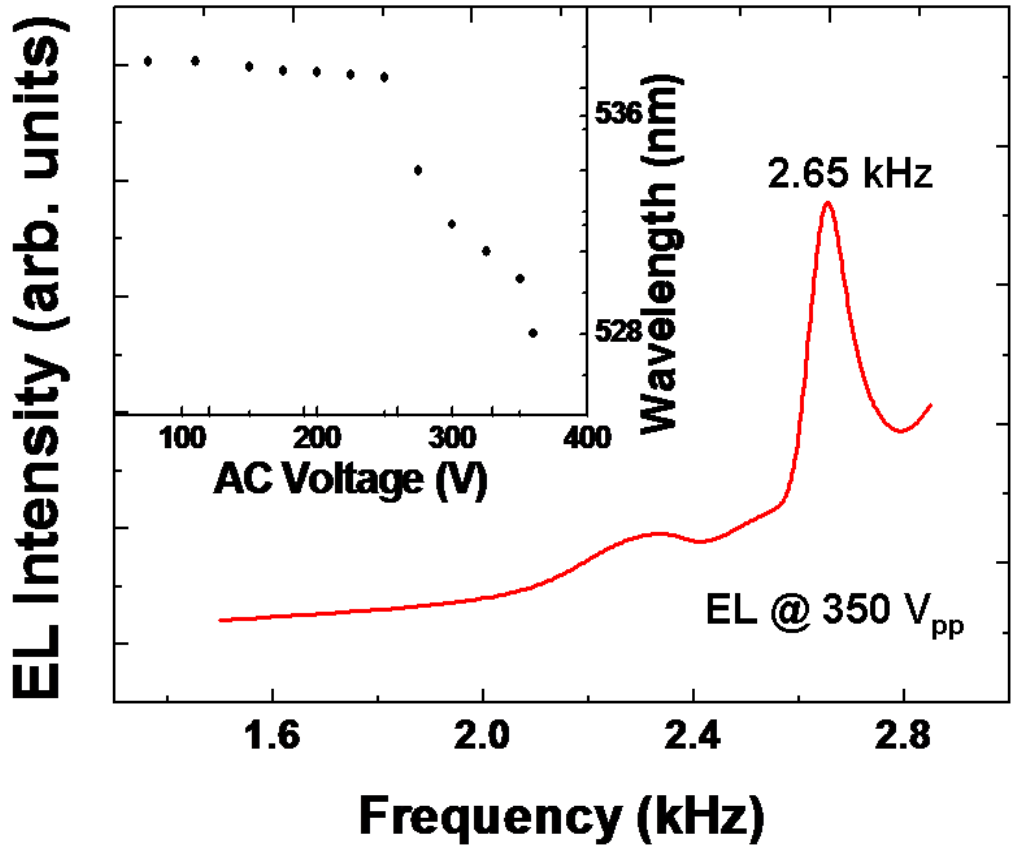

**Figure 7.** Plot of the EL response as a function of drive frequency at a fixed AC voltage 350 $V_{pp}$. The inset shows a marginal (~12 nm) variation of emission wavelength flaunted by ZnS/MWCNT hybrid material under various applied AC voltages.

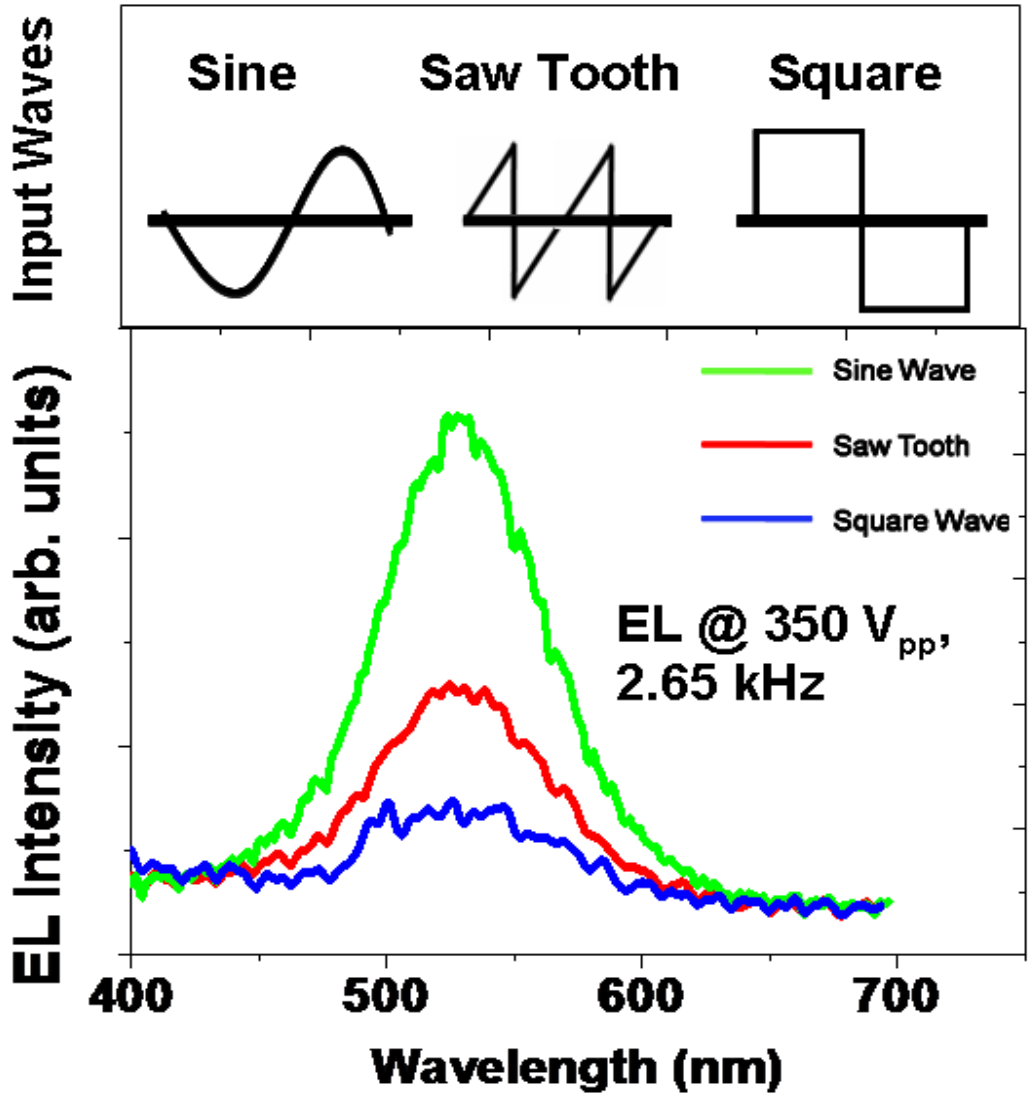

**Figure 8.** Plot of the EL response as a function of various input waveforms for a given bias conditions of 350 V$_{pp}$ at 2.65 kHz, which shows that the contribution of sine wave is optimum.

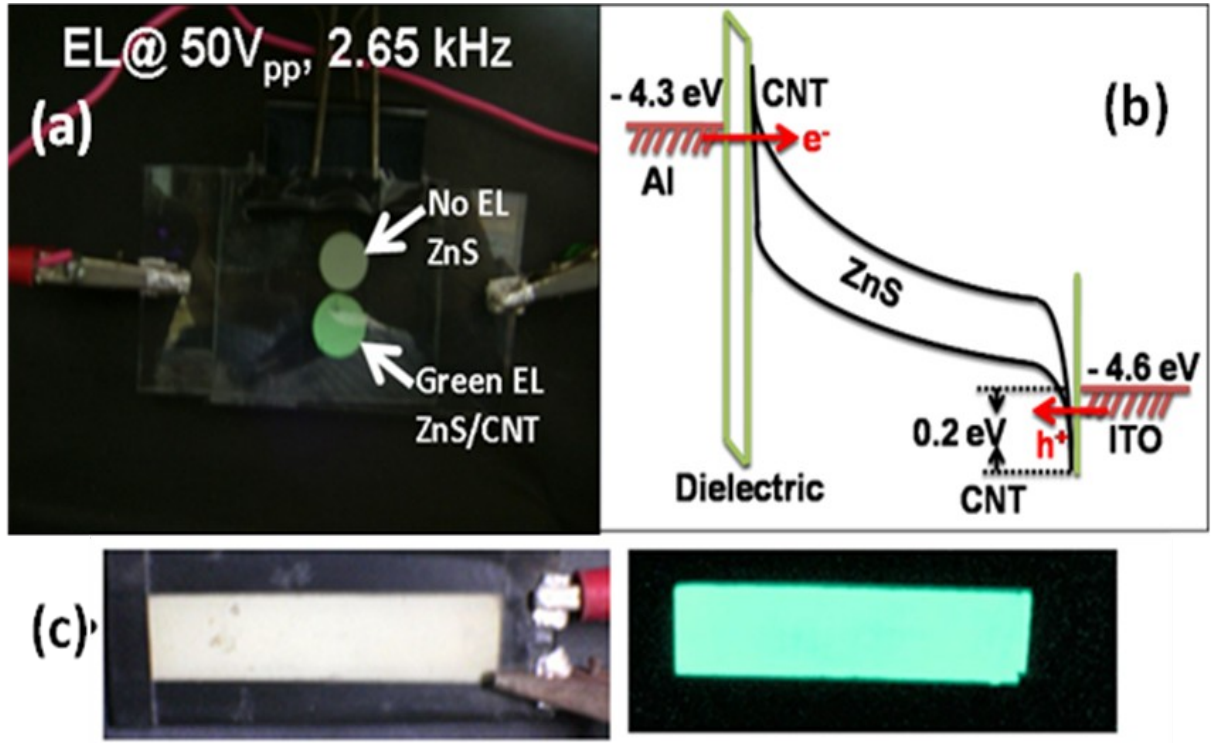

**Figure 9.** (a) A photograph of EL lamp having intrinsic ZnS:Cu phosphor and ZnS/MWCNT hybrid material at an applied AC voltage of 50 $V_{pp}$ at 2.65 kHz. The phosphor without MWCNT did not show EL even at applied AC voltages greater than 1000 $V_{pp}$, whereas, the ZnS/MWCNT hybrid material exhibited EL at threshold voltages as low as 50 $V_{pp}$ at 2.65 kHz. Both the materials are kept in a same lamp to compare the EL characteristics under similar conditions. (b) Role of CNTs in triggering the electroluminescence from ZnS:Cu phosphor. (c) The photographs of an actual EL lamp in room-light and dark conditions; exhibiting strong green EL emission operating at bias conditions of 350 $V_{pp}$ AC, 2.65 kHz and sine wave input.